\documentclass[aps,prl,superscriptaddress,reprint,floatfix,longbibliography,showpacs,amsfonts,amsmath,amssymb]{revtex4-1}
\DeclareUnicodeCharacter{2009}{\,}
\usepackage{multirow}
\usepackage{graphicx}
\usepackage{hyperref}
\usepackage{xr-hyper}
\usepackage{upgreek}
\usepackage{epstopdf}
\usepackage[note-name=, use-sort-key = false]{notes2bib}
\usepackage{color}
\usepackage{soul}
\usepackage{ulem}

    \hypersetup{
    	colorlinks = true,
		citecolor = blue,
		bookmarksnumbered = true,
		urlcolor = blue
	}

\begin{document}

\title{Magnetoelastic Signatures of the Conical State and Charge Density Waves in Antiferromagnetic FeGe}

\author{L. Prodan}
\email{lilian.prodan@uni-a.de}
\affiliation{Experimentalphysik V, Center for Electronic Correlations and Magnetism, Institute of Physics, University of Augsburg, D-86159 Augsburg, Germany}
\author{J. Sourd}
\affiliation{Hochfeld-Magnetlabor Dresden (HLD-EMFL) and Würzburg-Dresden Cluster of Excellence ctd.qmat,
Helmholtz-Zentrum Dresden-Rossendorf, 01328 Dresden, Germany}
\author{S. Zherlitsyn}
\affiliation{Hochfeld-Magnetlabor Dresden (HLD-EMFL) and Würzburg-Dresden Cluster of Excellence ctd.qmat,
Helmholtz-Zentrum Dresden-Rossendorf, 01328 Dresden, Germany}
\author{L. Chioncel}
\email{liviu.chioncel@uni-a.de}
\affiliation{Theoretische Physik III,  Institute of Physics, University of Augsburg, D-86135 Augsburg, Germany}
\affiliation{Augsburg Center for Innovative Technologies, University of Augsburg, 86135 Augsburg, Germany}

\begin{abstract}

Kagome systems host intertwined spin, charge, and lattice degrees of freedom that drive emergent collective states. Here, we identify two distinct energy scales in the noncollinear kagome magnet FeGe: a field-tunable magnetic fluctuation channel at $\sim 35$~K associated with the conical state, and a field-independent channel at $\sim 100$~K linked to charge-density-wave fluctuations. Ultrasound measurements provide direct access to the exchange-renormalized magnetic stiffness, whose softening governs the acoustic anomaly and follows a symmetry-constrained quadratic field dependence. We further predict a linear temperature dependence of neutron diffraction intensities at fixed field and a quadratic field suppression at fixed temperature. These results identify magnetic stiffness as a key control parameter of spin–lattice dynamics and establish ultrasound as a sensitive probe of coupled collective modes.
\end{abstract}

\maketitle
Kagome magnets provide a fertile platform for exploring emergent phenomena arising from competing interactions and coupled degrees of freedom~\cite{Balents2010, Zhao2020, Prodan2025, Subires2025}. In itinerant systems, the interplay of spin, charge, and lattice dynamics can generate a wide range of collective states, including nontrivial magnetism, charge ordering, and electronic instabilities~\cite{ Hirschberger2019, Altthaler2021, Yu2012, Ortiz2020, Kiyohara2016, Na2016, Prodan2023, Wilson2024, Prodan2024, Yin2018, Grandi2024, Jiang2021, Teng2022, Oh2025, Klemm2025}. Disentangling these intertwined degrees of freedom and identifying the underlying control parameters remain central challenges in correlated quantum materials.

\begin{figure}[h]
\centering
\includegraphics[width=0.45\textwidth]{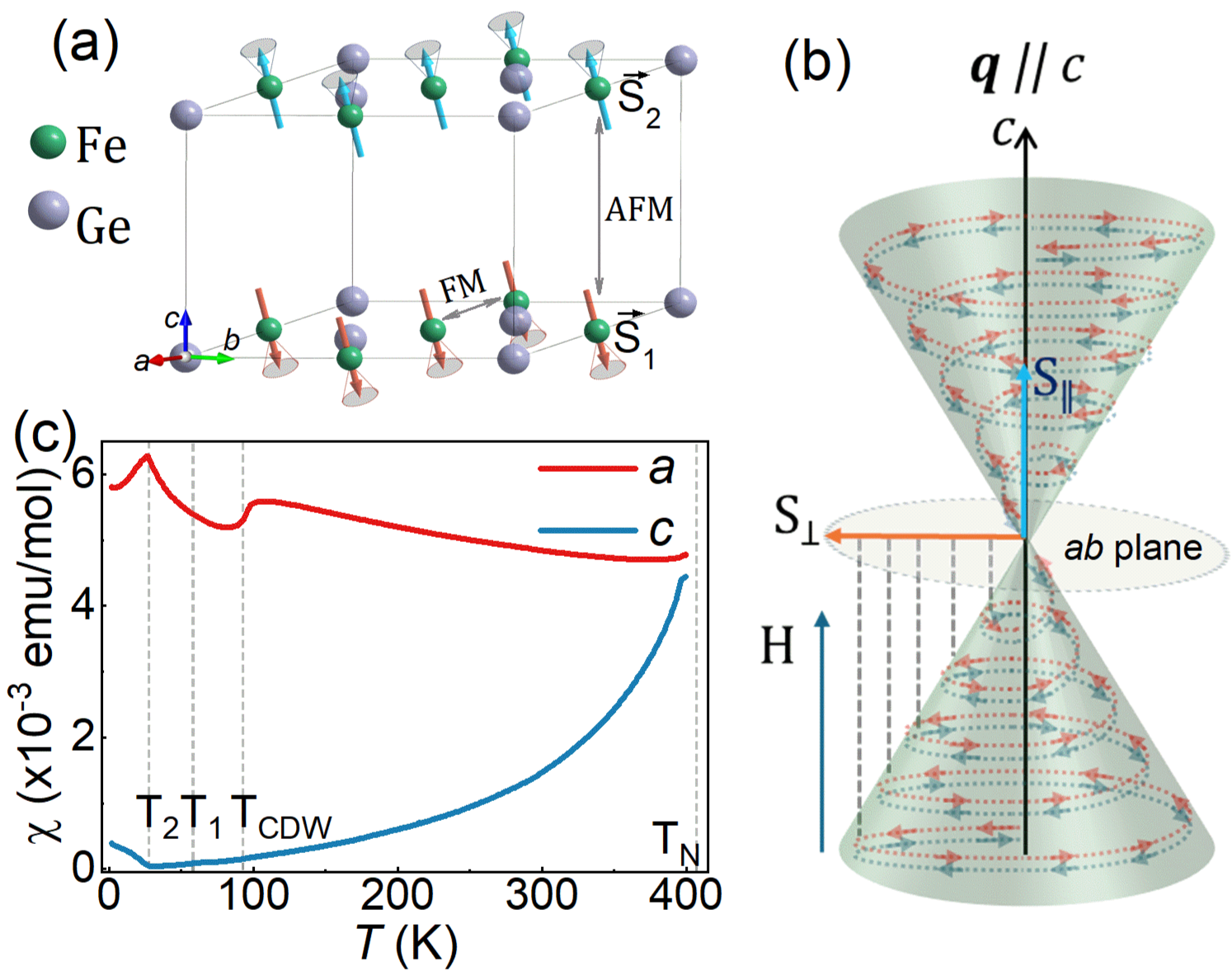}
\caption{(a) Crystal structure of hexagonal FeGe ($P6/mmm$) and the low-temperature magnetic spin arrangement. (b) Schematic view of the double-cone magnetic structure with propagation vector $\mathbf{q}\parallel c$ and its field evolution for $H\parallel c$; dashed lines indicate the field-induced change of the cone angle. (c) Temperature dependence of the magnetic susceptibility $\chi$ measured at 1~T for $H\parallel a$ and $c$.}
\label{fig:lcone}
\end{figure}

The hexagonal kagome antiferromagnet FeGe has recently emerged as a model system to study the intertwined degrees of freedom. It crystallizes in a centrosymmetric $P6/mmm$ structure, where Fe kagome layers are separated by Ge hexagonal networks [Fig.~\ref{fig:lcone}(a)]. Below $T_N \approx 410$~K, FeGe exhibits A-type collinear antiferromagnetic (AFM) order, with ferromagnetically aligned spins within the kagome planes and AFM stacking along the $c$ axis~\cite{Carrander1970, Bernhard1988}. In addition, a charge-density-wave (CDW) transition at $T_{\mathrm{CDW}} \approx 110$~K introduces a structural superlattice driven by partial dimerization of Ge atoms~\cite{Teng2022, Chen2022, Chen2024, Klemm2025, Wu2024, Shi2025, Bonetti2024, Wenzel2024}, accompanied by enhanced magnetic moment~\cite{Teng2022, Subires2025, Han2025, Zhou2023}. Upon cooling below $T_{\mathrm{1}} \approx 60$~K, the magnetic structure evolves into a canted double-cone state [Fig.~\ref{fig:lcone}(a),(b)], characterized by a finite in-plane spin component and a modulation along the $c$ axis~\cite{Bernhard1984,Bernhard1988,Beckman1973,Chen2024a,Klemm2025}. 

The coexistence of CDW order and noncollinear magnetism places FeGe in a regime where multiple collective modes interact on comparable energy scales. Low-energy spin, charge, and lattice fluctuations contribute to the relevant electronic susceptibilities, which renormalize the elastic constants \(C\) and thereby affect the propagation of acoustic waves with velocity \(v=\sqrt{C/\rho}\), where \(\rho\) is the mass density~\cite{Hauspurg2024, Luthi2007}. Consequently, sound velocity measurements provide direct access to fluctuation-induced changes in the elastic response. In metallic antiferromagnets, both magnons and low-energy spin fluctuations couple to strain via exchange-striction and magnetic anisotropy mechanisms, making the elastic constants sensitive to magnetic correlations and various competing ordered phases~\cite{Luthi2007,Zherlitsyn2014, Tsurkan2017,sourd2024}.

In this work, we performed sound velocity measurements to probe the exchange-renormalized magnetic stiffness of the conical state and its evolution under magnetic field. We establish a quantitative link between elastic softening and the transverse spiral component of the magnetic structure, consistent with neutron scattering. Furthermore, using a simplified renormalization-group framework (RG), we describe the effective coupling between magnetic and CDW fluctuations.
Our theoretical results demonstrate that the sound velocity variation provide direct access to coupled collective modes and identify magnetic stiffness as a central parameter governing spin–lattice dynamics in correlated quantum systems.

\begin{figure}[h!]
\centering
\includegraphics[width=0.42\textwidth]{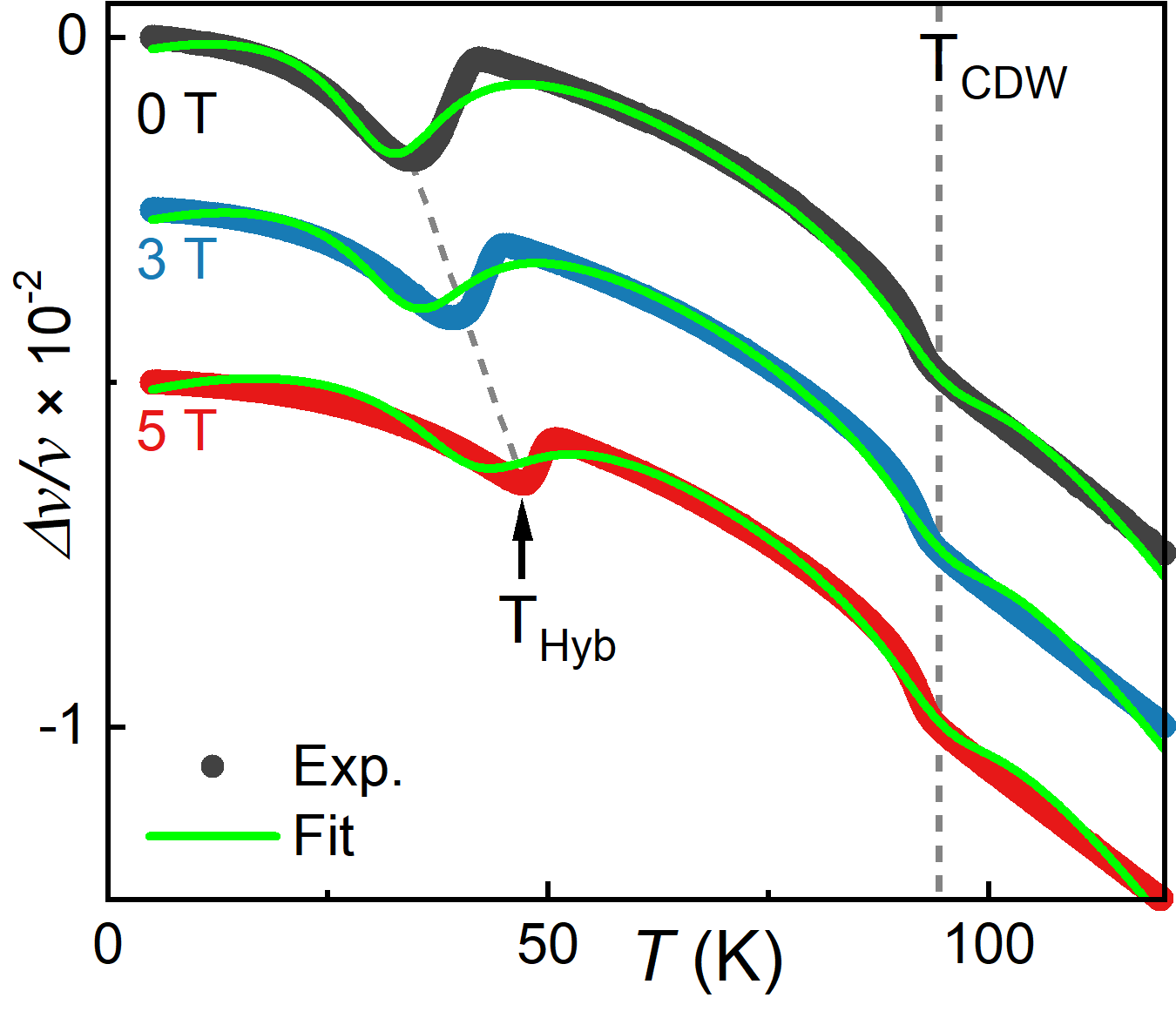}
\caption{Temperature dependence of the relative sound-velocity change, $\Delta v/v$, measured at 0, 3, and 5~T for the longitudinal acoustic mode ($\mathbf{k}\parallel c$, $\mathbf{u}\parallel c$) with $H\parallel c$. Dashed vertical lines indicate the CDW transition ($T_{\mathrm{CDW}}$) and the phonon–magnon hybridization temperature ($T_{\mathrm{Hyb}}$). The green line is a fit using Eq.~(\ref{eq:fit}). Curves at 3 and 5~T are vertically offset for clarity.}
\label{fig:latpar}
\end{figure}

High-quality single crystals of FeGe were grown by chemical vapor transport. Prior to ultrasound measurements, the samples were characterized by x-ray spectroscopy and magnetization. (For details see section I in Supplemental Material (SM)~\cite{SM}). Figure~\ref{fig:lcone}(c) shows magnetic
phase transitions characteristic of kagome FeGe~\cite{Teng2022,Bernhard1988}. 
The ultrasound measurements were performed utilizing the transmission pulse-echo technique with phase sensitive detection~\cite{Zherlitsyn2014, Hauspurg2024}. Figure~\ref{fig:latpar} shows the experimentally measured relative change in the sound velocity, $\Delta v/v$, for the longitudinal acoustic mode propagating along the $c$ axis. In zero magnetic field, $\Delta v/v$ exhibits pronounced anomalies near $\sim 35$~K and around $\sim 100$~K, coinciding with temperatures comparable to those associated with the onset of the canted conical magnetic state and the formation of the CDW phase [Fig.~\ref{fig:lcone}(c)]. Upon applying a magnetic field along the $c$ axis, the minimum associated with the conical state shifts to higher temperatures, reaching approximately $45$~K at $5$~T. The amplitude of the relative change $\Delta v/v$ decreases with increasing field, indicating a field-induced modification of the longitudinal acoustic mode. In contrast, the anomaly at $T_{\mathrm{CDW}}$ remains nearly field independent. 

To describe the coupled magnetoelastic response, we employ a multimode coupling framework~\cite{li.va.82}. Within the weak coupling approach, when phonons couple to magnons and CDW fluctuations, the phonon propagator becomes renormalized. The physical phonon frequencies follow from the poles of the dressed propagator. For the acoustic dispersion $\omega_q=vq$, assuming weak coupling, $\omega_{\mathbf q}\approx \omega^{(0)}_{\mathbf q}+\delta\omega_{\mathbf q}$, and expanding to leading order yields the relative velocity change: 
(see also SM, Sec. II~\cite{SM}):
\begin{align}
\frac{\Delta v}{v}
=
\frac{\Re \Pi^{\mathrm{ph}}(\mathbf q,\omega^{(0)}_{\mathbf q})}
{2\big(\omega^{(0)}_{\mathbf q}\big)^2} =
\frac{1}{2C_0}
\Re \Pi^{\mathrm{ph}}(\mathbf q\rightarrow0,\omega\rightarrow0),
\label{eq:dv/v}
\end{align}
where $C_0=\rho v_0^2$ is the bare elastic constant. 
In the long-wavelength limit ($\mathbf q\!\to\!0$) ultrasound probes the quasi-static elastic response.
The imaginary part of the self-energy governs sound absorption,
$\alpha_{\mathrm{att}} \propto {\Im \Pi^{\mathrm{ph}}(\mathbf q,\omega^{(0)}_{\mathbf q})}/{2\omega^{(0)}_{\mathbf q}}$.
Thus, all microscopic information enters through the renormalized phonon self-energy $\Pi^{\mathrm{ph}}$.
Within perturbation theory, the phonon self-energy can be expressed in terms of collective susceptibilities~\cite{li.va.82,be.gi.23}, and reads as:
\begin{align}\label{eq:Pi_chi}
\Pi^{\mathrm{ph}}(\mathbf q,\omega)
=
g_m^2\,\chi_m(\mathbf q,\omega)
+
g_c^2\,\chi_c(\mathbf q,\omega),
\end{align}
where $g_{m}$ and $g_{c}$ denote bilinear coupling constants between strain and the magnetic or CDW order-parameter fluctuations, respectively. This result can also be derived by Gaussian integration over the auxiliary fields and represents the leading-order self-energy correction generated by bilinear strain-order-parameter coupling (see SM, Sec. III~\cite{SM}).
In the long-wavelength limit, the relative sound-velocity change is therefore determined by the real part of the static self-energy, Eq.~\eqref{eq:dv/v}. 

In the following, rather than pursuing a direct microscopic computation, we parameterize many-body susceptibilities to describe the experimental data.
Susceptibilities can be computed using standard many-body procedures~\cite{maha.13} as demonstrated in the SM~\cite{SM}.
%
Their expressions provide a natural starting point for constructing a fit model for the sound-velocity renormalization, according to Eq.~\eqref{eq:dv/v} with a proper consideration of low frequency and momenta limits~\footnote{The relevant quantity is the real part of the phonon self-energy taken in the limit $q\rightarrow0$ and $\Omega_n\rightarrow0$. Since the order of limits may not commute, one should first take the long-wavelength limit $q\to0$ along the acoustic branch ($\omega\sim vq$) before sending $\Omega_n\to0$, corresponding to the experimental ultrasound condition.}.
Formally, the resulting static self-energy reads: 
\begin{align}
\Pi^{ph}(0,0) = 
\frac{g_m^2}
{\omega_0^2(T)+\alpha H^2+\Gamma^2}
\nonumber
+  \frac{g_c^2}
{a_c(T-T_{\mathrm{CDW}})^2+\Gamma_c^2}.
\label{eq:self-energ}
\end{align}
The magnetic contribution corresponds to a damped resonance originating from hybridization between acoustic phonons and transverse spin-wave modes. 
The CDW channel describes a second-order instability, yielding a Lorentzian-like temperature dependence.
In addition to fluctuation-induced effects, the sound velocity exhibits a smooth temperature dependence  arising from lattice anharmonicity and thermal expansion.
According to Eq.~\eqref{eq:dv/v}, the relative sound-velocity change can be decomposed as follows: 
\begin{align}
\frac{\Delta v}{v}
=
\left(\frac{\Delta v}{v}\right)_{\mathrm{bg}}
+
\left(\frac{\Delta v}{v}\right)_{\mathrm{mag}}
+
\left(\frac{\Delta v}{v}\right)_{\mathrm{CDW}} 
\end{align}
with:
\begin{align}
\left(\frac{\Delta v}{v}\right)_{\mathrm{bg}} = a_0+a_1T+a_2T^2.
\end{align}
where $a_0$ is the zero-temperature reference, $a_1$ accounts for linear thermal expansion, and $a_2$ captures the leading anharmonic contribution.
We model the magnetic contribution as a damped resonance,
\begin{equation}
\chi_m^{-1}(T,H) \propto
{\left[a_m(T,H)-\omega_0\right]^2 + \Gamma^2(H)} ,
\label{eq:magnetic}
\end{equation}
where the effective magnetic stiffness is parameterized as
$ a_m(T,H) = a_{m0} + bT + \alpha_m H^2$, the field-dependent damping as
$\Gamma(H) = \Gamma_0 + \gamma H^2$
and $\omega_0$ denotes the acoustic mode energy scale entering the hybridization condition. 
Physically, $a_m(T,H)$ represents the renormalized magnetic stiffness controlling the proximity to a magnetoelastic resonance with the acoustic branch. The quadratic field dependence is the lowest-order symmetry-allowed form in a centrosymmetric system; a negative $a_{m}$ corresponds to field-induced softening of the relevant magnetic mode.
In the denominator of $\chi_m(T,H)$, the parameters $a_{m0}$ and $\omega_0$ are not independent, as they enter only through their difference. We therefore define $\delta_0 = a_{m0} - \omega_0$, effectively absorbing $\omega_0$ into $a_{m0}$. The magnetic contribution then reads:
\begin{equation}
\left(\frac{\Delta v}{v}\right)_{\mathrm{mag}} = -
\frac{G_m}{\left[\delta_{0}+bT+\alpha_m H^2\right]^2
+ \left(\Gamma_0+\gamma H^2\right)^2}.
\end{equation}
where $G_m$ is the effective magnetoelastic coupling strength.

Near a continuous CDW transition, Gaussian critical fluctuations yield an Ornstein--Zernike form which reduces to a Lorentzian-like susceptibility in the static limit: 
\begin{equation}\label{eq:dv_v_cdw}
\left(\frac{\Delta v}{v}\right)_{\mathrm{CDW}}
= -\frac{G_c}{a_c (T - T_{\mathrm{CDW}})^2 + \Gamma_c^2}.
\end{equation}
Here $G_c$ is the CDW--elastic coupling strength and $\Gamma_c$ is the damping rate of the soft electronic mode approaching the CDW transition. The absence of a significant field dependence of this term in the fit supports its predominantly electronic origin and distinguishes it from the magnetically driven low-temperature anomaly. 
Both  $a_c$ and $\Gamma_c$ control the width of the velocity shift Eq.~\eqref{eq:dv_v_cdw}, 
and thus, one can absorb $a_c$ in $\Gamma_c$, i.e. in the fit procedure we can fix $a_c=1$ and fit only $\Gamma_c$.
The full fit function used for the global multi-mode analysis is:
\begin{align}
\frac{\Delta v}{v}(T,H) &= a_0 + a_1 T + a_2 T^2 \nonumber \\ 
&- \frac{G_m}{\left[\delta_{0}+bT+\alpha_m H^2 \right]^2 + \left(\Gamma_0+\gamma H^2\right)^2} \nonumber  \\
&- \frac{G_c}{(T - T_{\mathrm{CDW}})^2 + \Gamma_c^2}.
\label{eq:fit}
\end{align}

The simultaneous description of three $\Delta v/v$ curves, at $H=0, 3, 5$~T, with a single parameter set (see Table SM1) demonstrates that the decomposition into magnetic and CDW channels is internally consistent.

Figure~\ref{fig:latpar} shows that the global fit captures the pronounced low-temperature anomaly near 35~K, its suppression and evolution with magnetic field, and the nearly field-independent CDW shoulder around 100~K.
The minimum in the sound velocity occurs when the characteristic magnetic energy scale becomes comparable to the acoustic energy scale, $a_m(T,H) = \omega_0$, which yields the field-dependent dip position 
$T_{\mathrm{dip}}(H) = (\omega_0 - a_{m0} - \alpha_m H^2)/{b}$.
This condition indicates that the softening of the magnetic mode reduces its excitation energy toward the acoustic phonon energy, thereby maximizing magnetoelastic hybridization. As the magnetic field increases, the $ \alpha_m H^2 $ term shifts the softening condition to higher temperatures, resulting in a displacement of the anomaly.

\begin{figure}[h]
\centering
\includegraphics[width=0.40\textwidth]{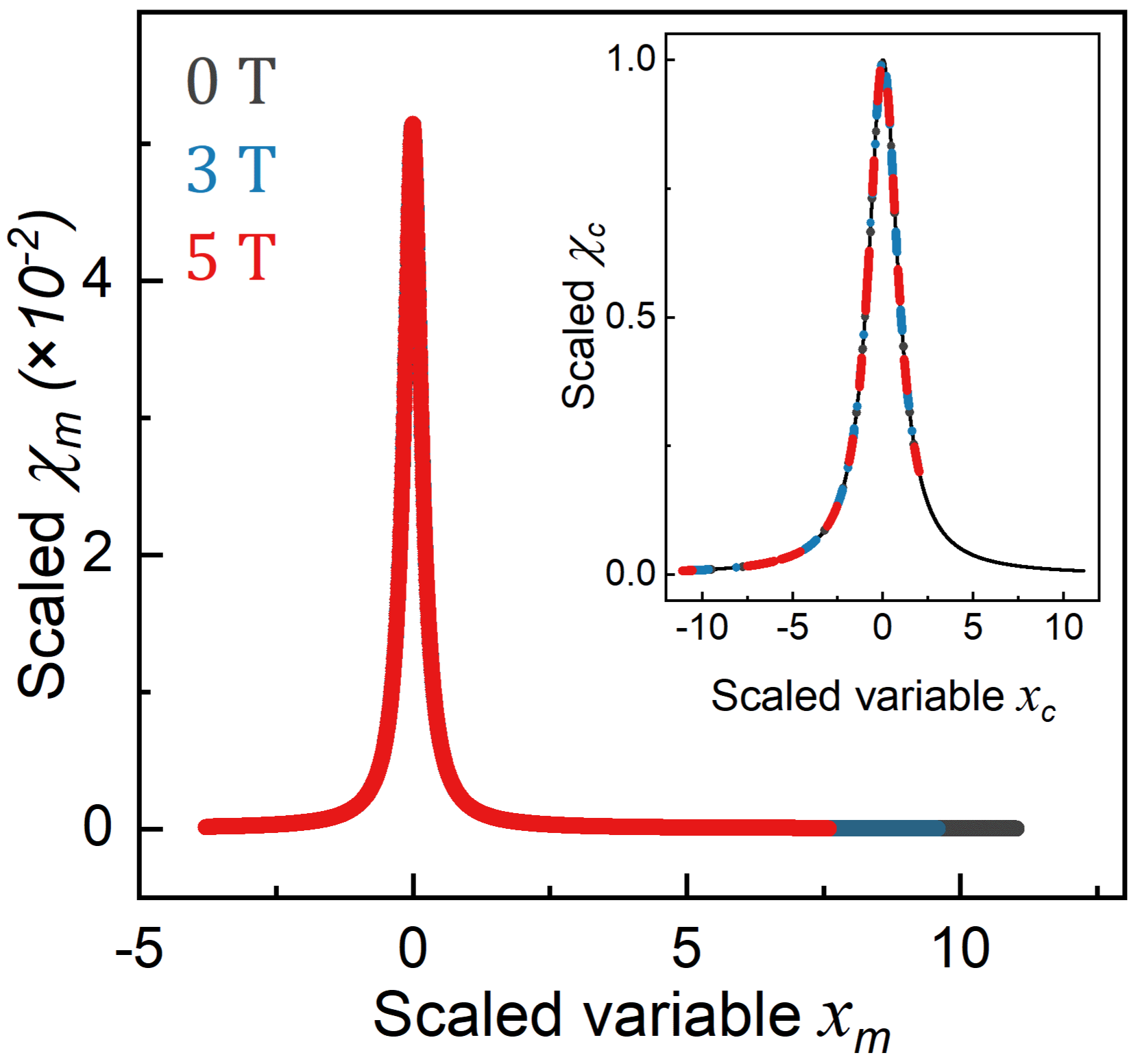} 
\caption{
Scaling collapse of magnetic and CDW (inset) susceptibilities.}
\label{fig:scale_m+cdw}
\end{figure}

While the multi-mode analysis decomposes the ultrasound-velocity anomalies into magnetic and CDW contributions, it does not by itself demonstrate that these components follow universal fluctuation forms. 
The scaling and collapse of mode susceptibilities (governed by critical or near-critical fluctuations) indicates, however, that the fitting parameters correspond to energy scales of intrinsic fluctuations of the system~\footnote{In the scaling analysis we neglect the smooth background contribution
modeled as a second-order polynomial in $T$ which represents analytic, non-critical lattice effects that are field independent. 
Since it does not contain intrinsic fluctuation energy scales, it does not participate in scaling behavior.}.
The magnetic contribution to the velocity shift written using dimensionless ratios of relevant energy scales has the form:
\begin{align}
\chi_m(T,H) = \frac{G_m}{\delta(T,H)^2 + \Gamma(T,H)^2}
           = \frac{G_m}{\Gamma(T,H)^2} f_m(x_m), \nonumber 
\end{align}
with $\delta(T,H) = \delta_0 + bT + \alpha_m H^2$ and 
$\Gamma(T,H) = \Gamma_0 + \gamma H^2$. 
Factoring out the damping scale introduces the dimensionless scaling variable $x_m = {\delta(T,H)}/{\Gamma(T,H)}$ and a corresponding Lorentzian function $f_m(x_m) = {1}/{(1 + x_m^2)}$.
After rescaling by the field-dependent amplitude $G_m/\Gamma(T,H)^2$, all magnetic data collapse onto a single Lorentzian curve (Fig.~\ref{fig:scale_m+cdw}).
Similarly, the CDW contribution is described by
\begin{equation}
\chi_c(T) = \frac{G_c}{(T-T_{\mathrm{CDW}})^2 + \Gamma_c^2}
           = \frac{G_c}{\Gamma_c^2} f_c(x_c),
           \label{eq:CDW}
\end{equation}
where $x_c = (T - T_{\mathrm{CDW}})/{\Gamma_c}$ and $f_c(x_c) = 1/(1 + x_c^2)$.
Here, $\Gamma_c$ sets the characteristic fluctuation energy scale. 
The observed collapse confirms that the CDW response depends only on the reduced distance from $T_{\mathrm{CDW}}$.
The separate Lorentzian forms indicate that magnetic and CDW fluctuations coexist over the same temperature range but retain distinct intrinsic fluctuation scales.
The peak amplitudes of the scaling functions are given by $\chi_c^{\mathrm{max}} = {G_c}/{\Gamma_c^2}$ and $\chi_m^{\mathrm{max}} = {G_m}/{\Gamma^2}$.
From our numerical analysis we find $\chi_c^{\mathrm{max}} \gg \chi_m^{\mathrm{max}}$.
Within an Ornstein-Zernike framework, where $\chi \sim \xi^2$, the larger CDW amplitude is consistent with more extended CDW correlations and/or stronger coupling to the elastic degrees of freedom.

To understand the evolution of the effective coupling between the magnetic and CDW phases, we interpret the scaling within a renormalization-group (RG) inspired technique~\cite{card.96}, in which the coexistence of these modes is an input at the level of the degrees of freedom, and possible fixed points (FP) are outcomes of the RG flow (see also SM, Sec. IV~\cite{SM}). Thus, the full scaling form for the velocity shift is written as:
\begin{align*}
\frac{\Delta v}{v} - 
\left(\frac{\Delta v}{v}\right)_{\mathrm{bg}}
 = -\frac{1}{\Gamma^2}\frac{1}{1 + x^2_m} -  \frac{1}{\Gamma^2_c} \frac{1}{1+x^2_c} \ . 
\end{align*}
Each scaling variable $x_m$ and $x_c$ define independent relevant directions associated with magnetic and CDW instabilities.   
The point (0,0) corresponds to the Gaussian fixed point of the linearized scaling problem, although it is not directly realized under the experimental conditions (see discussion in the SM, Sec. IV \cite{SM}).
\begin{figure}[h]
\centering
\includegraphics[width=0.40\textwidth]{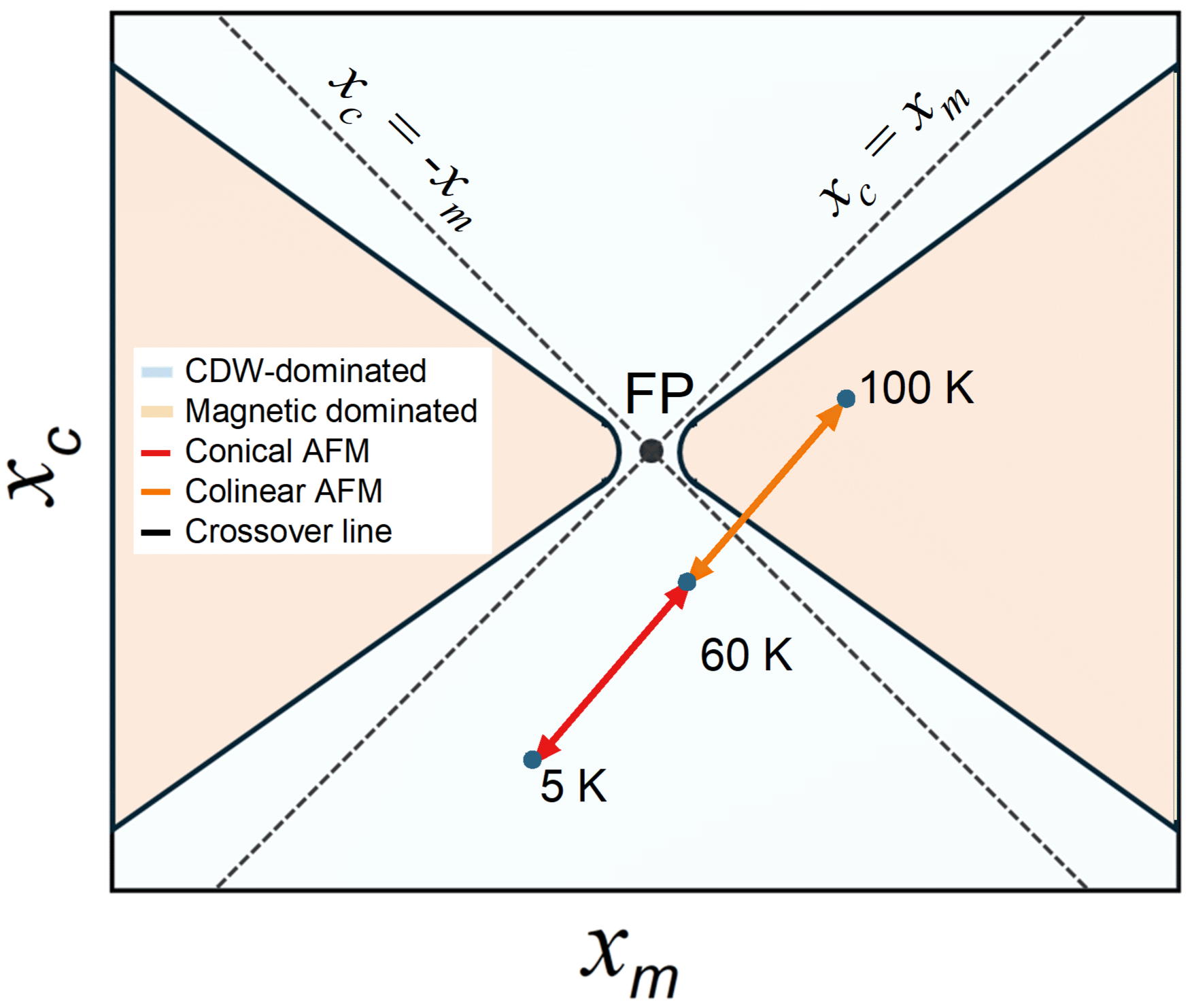} 
\caption{
Two-parameter scaling space spanned by $(x_m, x_c)$. 
Canted arrows indicate the thermodynamic trajectory of the system in the $(x_m, x_c)$ space between 5 and 100~K.
}
\label{fig:flow_eq}
\end{figure}
The diagram in Fig.~\ref{fig:flow_eq} illustrates the linear RG eigen-directions, $x_{\pm}=x_m \pm x_c$, corresponding to the separatrices of the flow. These directions coincide with the diagonals in the $(x_m,x_c)$ plane and represent the dominant and subdominant scaling directions in the special case $\Gamma=\Gamma_c$.
Moving along $x_m$ corresponds to tuning through the magnetic instability (e.g., cone opening or closing), whereas moving along $x_c$ corresponds to tuning through the CDW instability.
When $x_m \rightarrow 0$, the system becomes soft relative to its damping scale, which leads to the observed scaling collapse.
The crossover hyperbola (thick black line seen in Fig.~\ref{fig:flow_eq}) represents the points in the ($x_m,x_c$)-plane obeying the equation $x^2_m = 2.5 x^2_c +1.5$ with $\Gamma^2_c/\Gamma^2_0 \approx 2.5$.   
It results from matching the energy scale of the two phases, and marks where magnetic fluctuations are as strong as CDW fluctuations.
As $\Gamma_c/\Gamma_0 > 1$ the CDW fluctuations are dynamically broader, the magnetic mode is sharper, and the magnetic dominance region is smaller near the origin. 

The thermodynamic trajectory is illustrated in Fig.~\ref{fig:flow_eq} by the tilted red–orange line, which traces the evolution of the system between 5 and 100~K in the $(x_m, x_c)$ plane. Along this path, the dominant fluctuations change from magnetic at low temperatures to CDW at higher temperatures. 
Within the RG picture, the CDW energy scale remains parametrically large over the entire temperature range, while magnetic order (the conical phase) emerges within a robust CDW background. 
This interpretation is consistent with the observed horizontal shifts and scaling collapse of $\Delta v/v$ when approaching the CDW regime, and shows that FeGe resides in a mixed regime where both $x_m$ and $x_c$ remain finite.

Neutron diffraction experiments~\cite{Bernhard1984,Bernhard1988} characterize the magnetic cone structure in terms of the total ordered moment amplitude $S^2 = S_\parallel^2 + S_\perp^2$ and the cone half-angle $\theta = \tan^{-1} (S_\perp/S_\parallel)$.
For $\theta = 0$, the system reduces to the collinear AFM state, while $\theta > 0$ corresponds to the conical phase.
As the temperature decreases, the cone angle increases, leading to a growth of the transverse component $S_\perp$, while the interlayer pitch $q$ remains nearly unchanged.
To assess how close the system is to an instability or whether the magnetic mode is soft, we minimize the Landau free-energy density ($F$) with respect to the magnetic order parameter amplitude $S$,
\begin{equation}
F = \frac{1}{2}\delta(T,H) S^2 + \frac{u}{4} S^4; \quad  S^2(T,H) = -\frac{\delta(T,H)}{u} \nonumber
\end{equation}
where $\delta(T,H) < 0$ controls the stiffness (energy cost) of the 
fluctuations of the order parameter $S$. 
The coefficient $u$ multiplying the quartic term $S^4$ ensures the stability of the ordered state and sets the scale of the order parameter amplitude. 
Thus, the scaling parameter $\delta(T,H)=\delta_0 + b T + \alpha_m H^2$ extracted from ultrasound directly controls the square of the ordered moment amplitude. Using the parameters obtained in the global fit 
(see Table SM1) the condition $\delta(T,H)=0$ defines the temperature at which the magnetic amplitude softens as:
\begin{equation}
T_c(H) = -\frac{\delta_0}{b} - \frac{\alpha_m}{b} H^2
        \approx 32.1~\mathrm{K} + 0.38 H^2 \ .
\end{equation}
This is consistent with the experimentally observed minima in the temperature dependence of the characteristic $\Delta v / v$ under applied magnetic field.

Neutron diffraction~\cite{Bernhard1984,Bernhard1988} measures the intensity of magnetic satellites associated with the transverse component.
From the minimization criteria, the leading temperature and field dependence up to smooth geometric factors from field-induced changes of the cone angle is:
\begin{align}\label{eq:I_sat}
I_{\rm sat}(T,H) \propto S_\perp^2 =S^2 \sin^2\theta \propto -\delta(T,H) \ .
\end{align}
Using the fit form of $\delta(T,H)$, Eq.~\eqref{eq:I_sat} predicts a linear temperature dependence of $I_{\rm sat}$ at fixed field, and quadratic suppression with magnetic field at fixed temperature. The coefficient $\alpha_m = -0.266$ \textit{T}$^{-2}$ extracted from ultrasound directly quantifies the field sensitivity of the magnetic stiffness controlling the conical state.
Note that the term $\alpha_m H^2$ in $\delta(T,H)$ in the conical state modifies both 
$S_\parallel$, $S_\perp$, i.e., it changes the cone angle [see also Fig.\ref{fig:lcone}(b)]. Because the ordered moment amplitude satisfies $S^2 \propto -\delta$, the quadratic field dependence implies that the field effectively renormalizes the stiffness of the conical state. 
The scaling collapse of $\Delta v/v$ therefore indicates that the magnetoelastic anomaly is governed primarily by the amplitude of the ordered moment rather than by independent longitudinal and transverse fluctuations.

We therefore conclude that ultrasound probes the effective magnetic stiffness dynamically through magnetoelastic coupling, while neutron diffraction measures its static manifestation in the cone amplitude. 
The observed scaling collapse of $\Delta v/v$ thus supports a unified description in which the temperature and field evolution of the conical state are governed by a single exchange-renormalized stiffness parameter $\delta(T,H)$.

The magnetic scaling variable can be directly related to the cone angle of the exchange-driven double-cone structure determined by neutron diffraction. In the centrosymmetric $P6/mmm$ crystal structure, Dzyaloshinskii–Moriya interaction is forbidden, and the conical state arises from frustrated exchange. The scaling of $\Delta v/v$ therefore reflects the renormalization of an exchange-controlled magnetic stiffness rather than a chiral twisting mechanism. Since the neutron satellite intensity is proportional to the square of the transverse spiral component, and hence to the cone angle, the ultrasound softening directly tracks the evolution of the same transverse magnetic amplitude that produces the elastic Bragg satellites. This establishes a quantitative bridge between elastic and neutron probes: ultrasound measures the renormalized susceptibility of the same exchange-driven mode whose static amplitude is detected in diffraction.

In summary, we show that ultrasound velocity anomalies in FeGe arise from coupled magnetic and charge fluctuations that can be described within a unified magnetoelastic framework. The low-temperature anomaly originates from phonon–magnon hybridization governed by an exchange-renormalized magnetic stiffness, while the CDW contribution remains largely field independent, consistent with its electronic origin. The observed scaling establishes a direct link between elastic softening and the transverse component of the conical spin structure, providing a quantitative connection between ultrasound and neutron scattering probes. These results identify magnetic stiffness as the key parameter controlling spin–lattice dynamics in correlated quantum materials.

\begin{acknowledgments}
We acknowledge D. Vollhardt and I. Kézsmárki for helpful discussions and suggestions. This work was supported by the Deutsche Forschungsgemeinschaft (DFG) through TRR (Project No. 360 - 492547816), SFB 1143 (Project No. 247310070), and the Würzburg-Dresden Cluster of Excellence on Complexity, Topology and Dynamics in Quantum Matter – ctd.qmat (EXC2147, Project No. 390858490). We also acknowledge support from the HLD at HZDR, member of the European Magnetic Field Laboratory (EMFL).

\end{acknowledgments}

\bibliography{Refs-FeGe}

\end{document}


\title{Supplemental Material: Magnetoelastic Signatures of the Conical State and Charge Density Waves in Antiferromagnetic FeGe}

\author{L. Prodan}
\email{lilian.prodan@uni-a.de}
\affiliation{Experimentalphysik V, Center for Electronic Correlations and Magnetism, Institute of Physics, University of Augsburg, D-86159 Augsburg, Germany}
\author{J. Sourd}
\affiliation{Hochfeld-Magnetlabor Dresden (HLD-EMFL) and Würzburg-Dresden Cluster of Excellence ctd.qmat,
Helmholtz-Zentrum Dresden-Rossendorf, 01328 Dresden, Germany}
\author{S. Zherlitsyn}
\affiliation{Hochfeld-Magnetlabor Dresden (HLD-EMFL) and Würzburg-Dresden Cluster of Excellence ctd.qmat,
Helmholtz-Zentrum Dresden-Rossendorf, 01328 Dresden, Germany}
\author{L. Chioncel}
\email{liviu.chioncel@uni-a.de}
\affiliation{Theoretische Physik III,  Institute of Physics, University of Augsburg, D-86135 Augsburg, Germany}
\affiliation{Augsburg Center for Innovative Technologies, University of Augsburg, 86135 Augsburg, Germany}

\pacs{}
\maketitle
\section{Crystal growth and characterization}

Polycrystalline FeGe was prepared from high-purity elemental precursors: iron powder (99.99\%, ChemPur) and germanium pieces (99.999\%, ChemPur), weighed in a molar ratio of Fe:Ge = 1:1. The mixture was thoroughly ground, pressed into pellets, and sealed in an evacuated quartz ampoule. The ampoule was heated at 600\,$^\circ$C for $\sim 200$~hours to promote solid-state reaction and homogenization, followed by rapid quenching in cold water.

\begin{figure} 
\centering
     \includegraphics[scale=0.2]{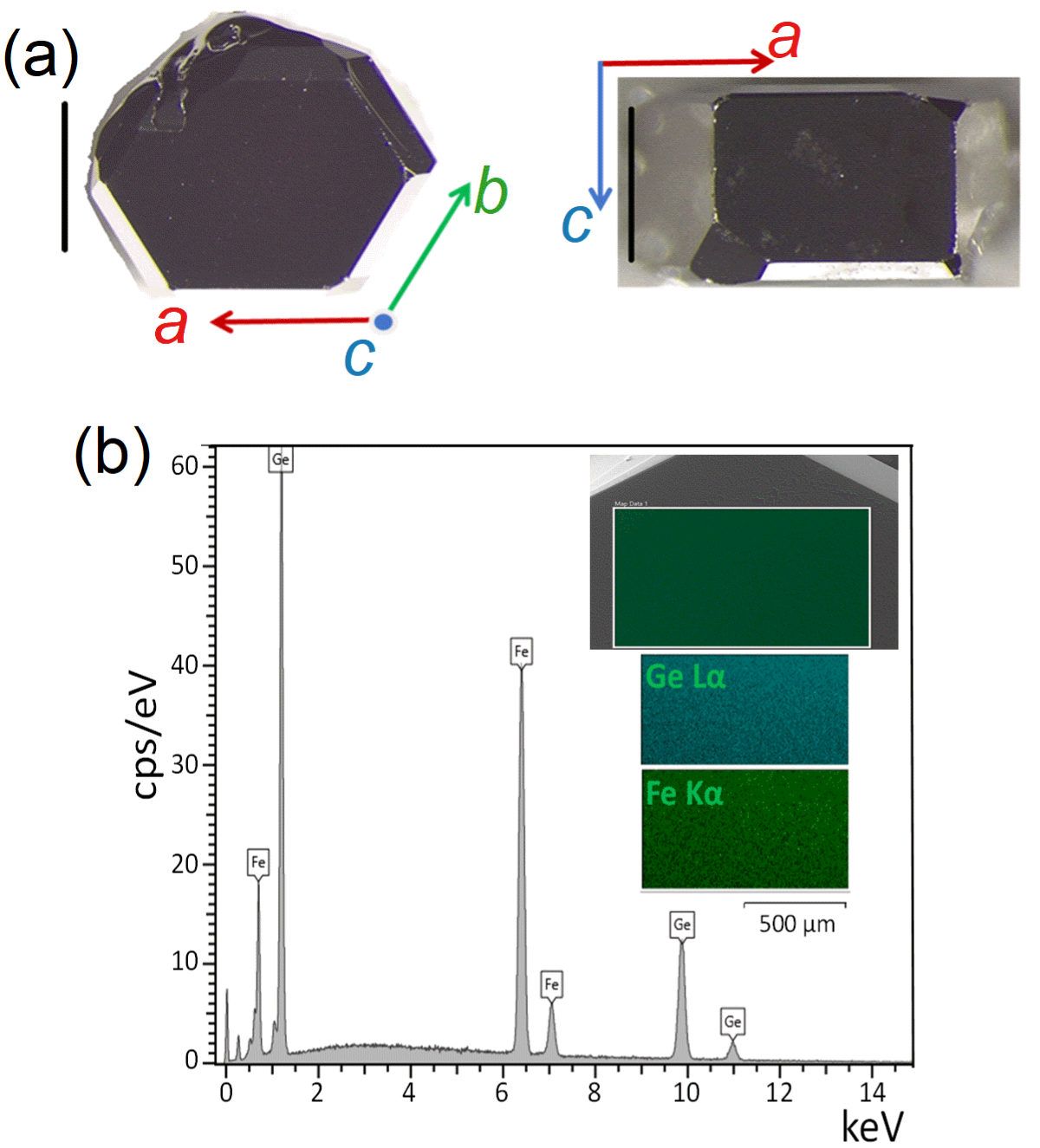}
     \caption{ Optical image of the as-grown FeGe. Vertical black bar corresponds to 1~mm size. Colored arrows indicate the hexagonal crystallographic directions. (b) Energy dispersive X-ray spectroscopy (EDS) spectra for single crystalline FeGe. The inset shows the scanning electron microscopy (SEM) image and elemental mapping for Fe and Ge atoms.  }
     \label{fig:figSM1}
 \end{figure}

Single crystals of FeGe were subsequently grown by the chemical transport reaction method. The polycrystalline precursor was ground and loaded into a quartz ampoule together with iodine (approximately 1:12 ratio) as the transport agent. The ampoule was evacuated to a pressure of $\sim 10^{-5}$~mbar and sealed. Crystal growth was carried out in a two-zone horizontal furnace, where the source zone was maintained at $\sim 520\,^{\circ}C$ with a temperature gradient along the ampoule of about $\Delta T = 50$\,$^\circ$C. After six weeks of transport, high-quality single crystals of FeGe, with well-defined crystallographic hexagonal $ab$ and rectangular $ac$ facets, approximately 1~mm$^3$ in size, were collected from the cold zone. An optical image of a FeGe single crystal with the as-grown facets is shown in Fig.~\ref{fig:figSM1}(a).

The phase purity, chemical composition, and homogeneity were verified by scanning electron microscopy (SEM) combined with energy-dispersive x-ray spectroscopy (EDS) using a ZEISS Crossbeam 550 system. EDS measurements were performed over multiple regions to ensure compositional homogeneity and confirmed the 1:1 stoichiometry. Figure~\ref{fig:figSM1}(b) shows the EDS spectrum of the crystal displayed in Fig.~\ref{fig:figSM1}(a) and used for magnetization measurements. The inset demonstrates the homogeneous spatial distribution of Fe and Ge across the sample surface.

Magnetic properties of the FeGe single crystals were studied using a SQUID magnetometer (Quantum Design MPMS 3). Field-dependent and temperature-dependent magnetization measurements were performed between 2 and 400~K under magnetic fields up to 7~T, applied both parallel and perpendicular to the crystallographic $c$-axis. These measurements were used to identify magnetic phase transitions characteristic of FeGe, Fig.1(c). The results confirm the antiferromagnetic state below 400~K, a transition near $T_{CDW}\approx 100$~K associated with the charge density wave (CDW) state~\cite{Teng2022}, and a spin reorientation transition below approximately $T_2\approx 30$~K~\cite{Bernhard1984, Bernhard1988}. No anomalies are observed in magnetic susceptibility at the onset of the canted double-cone state $T_1\approx 60$~K~\cite{Bernhard1984}.

The ultrasound measurements were performed on a different crystal from the same batch using a transmission pulse-echo technique with phase-sensitive detection~\cite{Zherlitsyn2014, Hauspurg2024}. For the longitudinal elastic modes both the propagation vector $k$ and
polarization $u$ were aligned along the $c$ axis. We used LiNbO$_3$ resonance transducers at fundamental frequency of 40 MHz. The experiments were performed in static magnetic fields using the Quantum Design PPMS system. 
\section{Relation between elastic constants and the renormalized phonon self-energy}
\label{sec:phonon_selfen}
To derive a microscopic description of the ultrasound velocity anomalies in a multi-component system such as FeGe --- involving acoustic phonons, longitudinal and transverse magnons, and CDW fluctuations --- we employ a Green's-function propagator formalism. Within this framework, phononic, magnonic, and CDW degrees of freedom are treated as coupled bosonic fields, and their interactions are described through retarded Green's functions that capture the dynamical response of the system.

For a bare longitudinal acoustic phonon the dispersion is given by
\begin{equation}
\omega^{(0)}_{\mathbf q}=v_0 q ,
\end{equation}
with bare sound velocity $v_0$. The corresponding retarded phonon Green's function is
\begin{equation}
D_{0}(\mathbf{q},\omega)
=
\frac{2\omega^{(0)}_{\mathbf q}}
{(\omega+i\eta)^2-\big(\omega^{(0)}_{\mathbf q}\big)^2}.
\end{equation}

When phonons couple to magnons and CDW fluctuations, the propagator becomes renormalized. The physical phonon frequencies follow from the poles of the dressed propagator,
\begin{equation}
\omega^2-\big(\omega^{(0)}_{\mathbf q}\big)^2
-
\Pi^{\mathrm{ph}}(\mathbf q,\omega)=0 ,
\end{equation}
where $\Pi^{\mathrm{ph}}(\mathbf q,\omega)$ denotes the effective phonon self-energy arising from magnonic and CDW channels.

Assuming weak coupling,
$\omega_{\mathbf q}\approx \omega^{(0)}_{\mathbf q}+\delta\omega_{\mathbf q}$,
and expanding to leading order yields
\begin{equation}
\delta\omega_{\mathbf q}
=
\frac{\Pi^{\mathrm{ph}}(\mathbf q,\omega^{(0)}_{\mathbf q})}
{2\omega^{(0)}_{\mathbf q}}
=
\frac{\Re \Pi^{\mathrm{ph}}(\mathbf q,\omega^{(0)}_{\mathbf q})}
{2\omega^{(0)}_{\mathbf q}}
+i
\frac{\Im \Pi^{\mathrm{ph}}(\mathbf q,\omega^{(0)}_{\mathbf q})}
{2\omega^{(0)}_{\mathbf q}} .
\end{equation}
The real part describes the frequency shift (sound-velocity renormalization), while the imaginary part determines the damping and ultrasound attenuation.

Using the acoustic dispersion $\omega_q=vq$, the relative velocity change becomes
\begin{align}
\frac{\Delta v}{v}
&=
\frac{\Re \Pi^{\mathrm{ph}}(\mathbf q,\omega^{(0)}_{\mathbf q})}
{2\big(\omega^{(0)}_{\mathbf q}\big)^2}
\nonumber \\
&=
\frac{1}{2C_0}
\Re \Pi^{\mathrm{ph}}(\mathbf q\rightarrow0,\omega\rightarrow0),
\label{eq:A_dv/v}
\end{align}
where $C_0=\rho v_0^2$ is the bare elastic constant. In this long-wavelength limit ($\mathbf q\!\to\!0$) ultrasound probes the quasi-static elastic response.

The imaginary part of the self-energy governs sound absorption,
\begin{equation}
\alpha_{\mathrm{att}} \propto
\frac{\Im \Pi^{\mathrm{ph}}(\mathbf q,\omega^{(0)}_{\mathbf q})}
{2\omega^{(0)}_{\mathbf q}} .
\end{equation}
Thus, all microscopic information enters through the renormalized phonon self-energy $\Pi^{\mathrm{ph}}$.

\section{Fitting formula and parameters}
\label{sec:fit_form}
The full fit function used for the global multi-mode analysis is:
\begin{align}
\frac{\Delta v}{v}(T,H) &= a_0 + a_1 T + a_2 T^2 \nonumber \\ 
&- \frac{G_m}{\left[\delta_{0}+bT+\alpha H^2 \right]^2 + \left(\Gamma_0+\gamma H^2\right)^2} \nonumber  \\
&- \frac{G_c}{(T - T_{\mathrm{CDW}})^2 + \Gamma_c^2}.
\label{eq:fit}
\end{align}

\begin{table}[h!]
\centering
\caption{Fit results with fixed $\Gamma_0 \approx 5.29$.
~\label{tab:I}}
\begin{tabular}{l c c c}
\hline
\textbf{Channel}& \textbf{param.} & \textbf{Value} & \textbf{Error} \\
\hline
\multirow{3}{*}{Background} 
& $a_0$ & $-0.1937 \times 10^{-3} $ & $4.0812\times 10^{-6}$ \\
& $a_1$ & $3.7233\times 10^{-5}$ & $1.7873\times 10^{-7}$ \\
& $a_2$ & $-8.3053\times 10^{-7}$ & $2.0109\times 10^{-9}$ \\
\hline
\multirow{5}{*}{Magnetic}
& $\delta_0$ & $-22.6542$ & $0.1721$ \\
& b & $0.70526$ & $0.00533$ \\
& $\alpha$ & $-0.26636$ & $0.00301$ \\
& $G_m$ & $0.05141$ & $0.00022$ \\
& $\gamma$ & $0.06052$ & $0.00109$ \\
\hline
\multirow{3}{*}{CDW}
& $G_c$ & $0.05911$ & $0.00245$ \\
& $T_\mathrm{CDW}$ & $98.245$ & $0.06283$ \\
& $\Gamma_c$ & $8.3818$ & $0.1528$ \\
\hline
\end{tabular}
\end{table}

Importantly, the same set of 12 fit parameters provides a simultaneous description of the temperature dependence of the relative sound-velocity change, $\Delta v/v$, measured at 0, 3, and 5~T, respectively.

\section{RG-flow analysis for FeGe}
\label{sec:RG_flow}

In FeGe, we consider a magnetic mode, a CDW mode, and a symmetry-allowed coupling $g$. The renormalization-group (RG) analysis addresses how these modes influence each other under a change of scale. 
The renormalization-group (RG) flow is described by the equations:
\begin{align*}
    \dot{x}_m = x_m + gx_c ; \quad 
    \dot{x}_c = x_c + gx_m 
\end{align*}
and are straight lines in the eigenbasis.
A fixed point satisfies: $\dot{x}_m=0$ and $\dot{x}_c=0$. In our linear model this occurs at $(x_m,x_c) =(0,0)$. 
The origin corresponds to the symmetric, disordered phase. 
The fixed point corresponds to $\delta(T,H) =0$ and $T=T_{CDW}$, which physically means that the magnetic mode is soft and the CDW mode is critical.
Diagonalizing the flow matrix yields the eigenvectors $v_+=(1,1)$ and $v_-=(1,-1)$ and correspond to the in-phase combination ($x_+=x_m+x_c$ magnetic + CDW) and out-of-phase combination 
($x_{-}=x_m-x_c$ 
magnetic - CDW). In our linearized weak-coupling flow, the origin is just the Gaussian fixed point.
Because the flow equations are linear and symmetric the only fixed point is the origin.
The linearized flow moves away from the origin along the eigen-directions, indicating which fluctuation channel dominates in a given region of the scaling plane rather than establishing true multicritical coexistence.
We derive the crossover line in the $(x_m, x_c)$ plane separating magnetic-dominated and CDW-dominated regimes using the scaled variables.
The crossover occurs when $\chi_m = \chi_c$ which leads to the following equation:
\begin{align*}
    x^2_m = \frac{\Gamma^2_c}{\Gamma^2}(x^2_c+1)-1 
\end{align*}
representing the equation of a hyperbola. When ${\Gamma^2_c}/{\Gamma^2} >1$ the hyperbola opens along $x_m$ while  ${\Gamma^2_c}/{\Gamma^2} < 1$ hyperbola opens along $x_c$. The special case  $\Gamma_c = \Gamma$ reduces to equal scales $|x_m| = |x_c|$ which geometrically represents the diagonal boundaries. 
The hyperbolas delimit the physical dominance regions.

We can extract an effective weak coupling $g_{eff}$ (entering the flow equations) from the experimental $\Delta v/v$ trajectories by projecting them onto the eigen-directions of the coupled RG flow. It amounts in computing the slope of the experimental trajectory ($\Delta x_c/\Delta x_m$).
From the fitted values of FeGe, $\Gamma_0 = 5.29$ and $\Gamma_c = 8.38$, and the ratio of the two is $\approx 2.5$ (numerical value taken from Tab.~\ref{tab:I}). 
Thus, the crossover equation is:
\begin{align*}
    x^2_m = 2.5 x^2_c +1.5 
\end{align*}
Solving the coupled RG equations we obtain the experimental trajectories that follow almost straight line, Fig.4, with slopes that slightly change with field.
The table below shows the characteristics of the experimental trajectory in the $(x_m,x_c)$ plane.

\begin{table}[h!]
\centering
\caption{Parameters of the experimental trajectory points in the $(x_m,x_c)$-plane at an external field of 5 T. Numerical values taken from Tab.~\ref{tab:I} 
}
\begin{tabular}{c c c c c c }
\hline
\textbf{T}& $x_m$  & $x_c$ & \textbf{Quadrant} & CDW & Dominant\\
\textbf{(K)}&  &  &  & order & channel\\
\hline
5~K    & -3.79 &  -11.13  & III & yes & CDW strong\\
60~K   & -1.85 &  -4.67  & III & yes &  CDW decreasing\\
100~K  &  2.02 &   0.94  & I & yes & mag. stiff\\
150~K  &  7.60 &   2.00  & I & no & mag. stiff\\
\hline
\end{tabular}
\end{table}

The trajectory arrow in Fig.4 of the main text points from low $T$ in the direction of high $T$. Along the path the magnetic structure changes from conical (III-quadrant) to AFM collinear (I - quadrant).
In the scaling units low $T$ ($x_m <0, x_c <0$ and $|x_m| < |x_c|$) is CDW dominant, but magnetism is present and conical, at high $T$ ($x_m >0, x_c >0$ and $|x_m| > |x_c|$) magnetic stiffness dominates, collinear AFM coexists with CDW, but CDW order is weaker.
Crossing the separatrix is not a phase transition; rather, it reflects a change in the relative weighting of the two scaling variables.
The trajectory curvature and slope gives a small ($g_{eff}\approx 0.071$) effective coupling between CDW and magnetic fluctuations.

The scaling behavior further suggests that the magnetic and CDW channels are only weakly coupled in the investigated field range, with small but measurable cross-coupling that bends the scaling trajectories in the $(x_m,x_c)$ plane. This weak mixing explains why the CDW shoulder remains largely field independent while the magnetic dip shifts systematically with field. In this sense, FeGe provides an experimentally clean realization of a two-channel magnetoelastic system.

\bibliography{Refs-FeGe}